# A High-fidelity Ensemble Simulation Framework for Interrogating Wildland-fire Behavior and Benchmarking Machine Learning Models

Qing Wang[1,a], Matthias Ihme[a,b,c], Cenk Gazen[a], Yi-Fan Chen[a], John Anderson[a]

[a]*Google, 1600 Amphitheatre Parkway, Mountain View, CA 94043, USA*
[b]*Department of Mechanical Engineering, Stanford University, Stanford, CA 94305, USA*
[c]*Department of Photon Science, SLAC National Accelerator Laboratory, Menlo Park, CA 94025, USA*

---

**Background**. Wildfire research uses ensemble methods to analyze fire behaviors and assess uncertainties. Nonetheless, current research methods are either confined to simple models or complex simulations with limitations. Modern computing tools could allow for efficient, high-fidelity ensemble simulations. **Aims**. This study proposes a high-fidelity ensemble wildfire simulation framework for studying wildfire behavior, assessing fire risks, analyzing uncertainties, and training machine learning (ML) models. **Methods**. We present a simulation framework that integrates the SWIRL-FIRE large-eddy simulation tool for wildfire predictions with the VIZIER optimization platform for automated run-time management of ensemble simulations and large-scale batch processing. All simulations are executed on tensor-processing units to enhance computational efficiency. **Key results**. A dataset of 117 simulations is created, each with 1.35 billion mesh points. The simulations are compared to existing experimental data and show good agreement in terms of fire rate of spread. Analysis is performed for fire acceleration, mean rate of spread, and fireline intensity. **Conclusions**. Strong coupling between wind speed and slope is observed for fire-spread rate and intermittency. A critical Froude number that delineates fires from plume-dominated to wind-dominated is identified and confirmed with literature observations. **Implications**. The ensemble simulation framework is efficient in facilitating large-scale parametric wildfire studies.

---

[1] Corresponding author
*Email address:* wqing@google.com (Qing Wang)

## Introduction

Wildfires pose serious threats to society, environment, and ecosystems as they can disrupt, damage, and destroy infrastructure, services, and properties (Barbero et al. 2015, Thomas et al. 2017). These threats are expected to grow with increasing severity and frequency of wildfires due to climate-pattern change, fire management, and the expansion of the wildland-urban interface (Abatzoglou & Williams 2016, Yoon et al. 2015).

While significant progress has been made on the fundamental understanding and modeling of wildfires, main challenges towards their accurate prediction, however, are the lack of reliable physical models and the consideration of their stochastic behavior (Riley & Thompson 2017). In particular, the occurrence, dynamics, and intensity of wildfires are affected by several factors, including variations in wind and vegetation, the primary ignition source resulting from natural or artificial causes, the local coupling of the fire with the environment, and long-range ignition by ember transport and spotting (Alexander & Cruz 2013, Ganteaume et al. 2013). In turn, uncertainties in these models arise from the lack of detailed understanding about the microphysics of the fuel consumption, the representation of computationally unresolved processes that involve the coupling between turbulence, chemistry, and radiation, as well as uncertainties from numerical approximations, spatio-temporal discretization, and boundary conditions (Cruz & Alexander 2013, Valero et al. 2021).

Ensemble methods have been developed to estimate uncertainties in physical models, variations in initial conditions, and inaccuracies of numerical models (Leutbecher & Palmer 2008, Wu & Levinson 2021). These methods are commonly employed in weather forecasting (Lewis 2005) and climate simulations (Giorgi & Francisco 2000, Murphy et al. 2004) to obtain probabilistic representations of the dynamics and response to uncertainties in physical parameters. Ensemble simulations have also been employed to examine fire-spread behavior affected by wind, fuel properties, topology, and ignition location (Anderson et al. 2005, Cruz 2010, Finney et al. 2011, Benali et al. 2016, 2017, Allaire et al. 2020) and data assimilation to integrate observations into simulations for fire-spread estimations (Mandel et al. 2008, Rochoux et al. 2014). In addition, ensemble methods have been employed to evaluate fire risks and subsequently optimize fire management strategies (Pinto et al. 2016). Given the high computational cost associated with conducting ensemble simulations, the majority of these investigations have utilized empirical models or 2D fire spread simulations (Finney et al. 2011), such as the Burn-P3 simulation model (Parisien et al. 2005), FlamMap (Finney 2006), and FarSite (Finney 1998). While several efforts have been made using high-fidelity physical simulations (Atchley et al. 2021, Pimont et al. 2012, Valero et al. 2021), the ensemble size in these cases has been relatively small due to computational limitations (Linn et al. 2007, Moinuddin et al. 2018).

Recent advances in computing hardware architectures, programming algorithms as well as data-driven methods and machine learning (ML) techniques (Ihme et al. 2022) offer opportunities to address these challenges. The objectives of this work are to develop and present a high-fidelity ensemble simulation framework for enabling the parametric examination of wildfire behavior, the creation of ensemble data to support ML training tasks, and to assist fire-risk assessments and uncertainty analyses of wildfires. The proposed simulation framework combines the open-source large-eddy simulation tool SWIRL-FIRE for wildfire predictions (Wang et al. 2023), an open-source optimization platform VIZIER (Golovin et al. 2017) for run-time management of ensemble simulations, and large-scale batch processing. The resulting framework is employed to perform detailed parametric analyses of the effects of changing wind and slope on the fire-spread behavior. Specifically, these two confounding factors have been identified as important parameters affecting the fire-spread behavior (Viegas 2004a), yet the detailed understanding of their dependencies on the fire behavior introduces significant experimental challenges due to the scale requirement to capture the coupling between the fire and atmospheric-flow environment (Clements & Seto 2015).

Experiments conducted on laboratory-scale fires have been used to investigate the influence of these two factors on fire statistics. Weise & Biging (1996) performed systematic experiments for changing wind speed and slope angle, and observed a strong impact of slope on the flame angle. The impact of the slope on the rate of spread (ROS) is not fully captured in several of the existing fire-spread models. Butler et al. (2007) experimentally studied the impact of slope on the ROS, considering six different slope angles with various fuel loadings and packing ratios. Depending on the slope angle, they identified three burning regimes in terms of flame attachment and heat-transfer mode. Silvani et al. (2012) showed that the slope angle not only changes the flame topology, but also the heat transfers ahead of the fire, resulting in the transition between different spreading regimes. Similar conclusions were obtained by Xie et al. (2017), where flame attachment was observed when the slope angle exceeds $25^\circ$ and convective heating become the dominating factor for preheating of unburned fuel instead of radiation compared to low slope angles. The impact of the slope angle on the convective heat transfer was further assessed by Liu et al. (2015). With experiments of nine different slope angles, they observed that the dominating convection mode transitions from natural to fire induced as the angle increases. The change of fire topology and transition between radiative and convective heating modes with respect to slope angle was confirmed experimentally and computationally by Sánchez-Monroy et al. (2019). Eftekharian et al. (2019) numerically investigated the impact of slope on the wind enhancement, concluding greater wind enhancement with positive slope due to fire-induced reverse pressure gradient by buoyancy.

While these studies have provided valuable information about the impact of each individual factor on the fire behavior, the combined effect of wind and slope remain only partially understood due to the scarcity of experimental data (Nelson 2002, Viegas 2004*b*, Weise & Biging 1994, 1996). Computational studies utilizing physical models have been conducted to explore these two factors individually (Innocent et al. 2023, Linn et al. 2010) and in combination (Pimont et al. 2012). Due to the high computational cost and complexity associated with these simulations, a comprehensive dataset to support the fundamental scientific analysis and evaluation of fire-spread models is currently not available.

In the present investigation, an ensemble dataset (FIREBENCH) is generated using SWIRL-FIRE. All simulations are performed on tensor-processing units (TPUs) (Jouppi et al. 2021) for computational efficiency. With this methodology, both fire and atmospheric dynamics are represented by physical models. Each case in this dataset encapsulates a time series of all flow field variables within the three-dimensional domain. The simulation framework is integrated with the open-source optimization platform VIZIER to facilitate automated sampling.

The remainder of this manuscript has the following structure. In the Methodology section, we present the methods employed to generate the ensemble simulation dataset, describing the numerical algorithm of SWIRL-FIRE and the framework for automated parameter sampling. In the Computational Setup section, we present the physical configuration, the parameter space, and sampling strategy employed in the present study. The results are presented in Results section, with specific focus on the quantitative analysis and detailed examination of the regime transition between plume-dominated and wind-dominated fires. In the Conclusions section we summarize the main findings and discuss potential utilities of the ensemble simulation framework and FIREBENCH dataset presented in this work.

## Methodology

This section presents detailed information about the numerical method, computational framework, and sampling technique employed in generating the FIREBENCH ensemble simulation dataset. Facilitated by the TPU architecture, we were able to conduct 117 high-fidelity ensemble simulations on a domain of 1500×250×1800 $m^3$ that is discretized with 1.35 billion grid points, resulting in 1.36 PiB of data that is efficiently stored on the Google Cloud platform, which is accessible at https://console.cloud.google.com/storage/browser/firebench.

### *Numerical Methods*

The simulations presented in this work are performed with SWIRL-FIRE (Wang et al. 2023), which is an open-source large-eddy simulation framework with a fully coupled combustion model to describe the interaction with the atmosphere. In this framework, the gas phase is described by the Favre-filtered conservation equations for mass, momentum, oxygen mass fraction, and potential temperature. The combustion of the solid fuel is modeled using a one-step mixing-limited oxidation reaction, considering the energetic impact of moisture evaporation (Linn et al. 2002), radiative losses, and convective heat transfer between fire and ambience. With this approach, physico-chemical processes involving pyrolysis, vaporization, and combustion are simplified following an assumption that the energy release is confined to regions where solid fuel is present (Linn 1997). The governing equations are solved with a low-Mach number approach (Wang et al. 2022) to alleviate limitations of the time-step size due to acoustic waves. For further details on the physical models, governing equations, and numerical algorithm, the reader is referred to Wang et al. (2023).

The solver was validated against the Fire-Flux II measurements (Clements et al. 2019), which considered the fire-spread behavior on a flat surface and prescribed wind profile. To extend this validation to conditions of interest to this study, we performed additional comparisons against the experiments by Morandini & Silvani (2010), considering different slope angles, wind conditions (speed and direction), fuel densities, and moisture content. Results from these simulations and comparisons with experimental data are presented in the Appendix of this study.

To consider complex terrains, we utilize the immersed-boundary (IB) method (Zhang & Zheng 2007) with direct forcing. This method provides a second-order accurate representation of the terrain without the necessity for unstructured meshes or intricate coordinate transformations, thereby retaining the overall efficiency of a structured mesh representation.

The SWIRL-FIRE simulation framework was implemented in TensorFlow (Abadi et al. 2015) using a just-in-time compilation approach to generate a highly optimized executable that fully exploits the hardware architecture for parallel execution on TPUs. The program is compiled using the Accelerated Linear Algebra (XLA) compiler, which generates a data flow graph with optimized computational and parallel efficiency. Linear weak and strong scalabilities have been demonstrated on canonical flows, showing its effectiveness for large-scale problems (Wang et al. 2022).

All simulations are performed on the fifth generation TPU v5e computing architecture (Jouppi et al. 2020). Each chip in this computing architecture consists of one tensor compute core, with four matrix-multiplication units to simultaneously perform four matrix-matrix multiplications of size $128^2$ on each core, providing a peak throughput of 197 teraflops. The TPU v5e chips are connected through a 2D Torus, with an inter-chip interconnect bandwidth of 1600 Gbps, forming a so-called TPU pod with 128 chips. The TPU v5e architecture inherently facilitates 8-bit arithmetic operations, while high-precision arithmetic operations can be achieved through software emulation. To achieve a balance between computational efficiency and numerical precision, simulations in this study are conducted using single precision.

Supported by this novel hardware architecture, we are able to perform simulations at significantly larger scales and with reduced time and energy utilization compared to those conducted with conventional CPU architectures. In this investigation, simulations on meshes of 1.35 billion mesh points are executed on 128 TPU chips and take approximately 13 hours of wall-clock time per simulation, resulting in a total of 1,664 TPU hours to create the FIREBENCH dataset.

*Ensemble Simulation Framework*

To enable high-fidelity ensemble simulations, we are leveraging the open-source optimization platform VIZIER (Golovin et al. 2017). Given only a parameter space and a cost function over this space, VIZIER efficiently searches for optimal parameters. In addition to providing various optimization algorithms, VIZIER is a system for managing compute-heavy jobs on multiple machines. In the present study, we are exploring the two-dimensional parameter space, consisting of wind speed and slope angle, as uniformly as possible but also taking advantage of the job management capabilities that VIZIER provides. To achieve this, we use a cost function that provides no information about the search space to VIZIER, thus forcing it to explore the parameter space uniformly.

VIZIER batch ensemble simulations are launched as an automated service through a Remote Procedure Call interface (Tay & Ananda 1990). After a simulation request is submitted, VIZIER will schedule the simulation runs automatically following configurations specified in the protocol buffer. Given the availability of the computational resources, VIZIER distributes runs across multiple machines to execute simulations in parallel. During these runs, the parameter sets being scheduled can be inspected dynamically through a dashboard interface.

In this study, a total of 117 individual ensemble simulations are scheduled with VIZIER, and are performed in parallel with ten simulations running simultaneously. *In situ* postprocessing scripts are executed by VIZIER to provide statistical analysis of the simulations, including 2D inspections of flow-field results and extraction of fire-front location, burn area, and flame length without manual intervention.

## Computational Setup

The configuration considered in this study is illustrated in Fig. 1 and consists of a 3D domain with size 1500×250×1800 $m^3$ along the streamwise ($x$), lateral ($y$), and vertical ($z$) direction, respectively. The computational domain is discretized with a structured Cartesian mesh with mesh size of 1.0×1.0×0.5 $m^3$, resulting in a total of 1.35 billion grid points. An Immersed Boundary (IB) method is employed to represent the terrain with constant slope angle $\alpha$. The slope starts at $x = x_0 = 100$ m downstream of the inlet and extends 1000 m along the horizontal direction. At the end of the slope, the terrain plateaus to provide well-described outflow conditions at the exit of the domain. The terrain is homogeneous along the lateral direction, and only slopes with positive angles are considered in the present study. For subsequent analysis, we introduce the $(\xi, \eta)$ coordinate system that is defined with respect to the slope distance $\xi = (x - x_0)/\cos(\alpha)$, see Fig. 1.

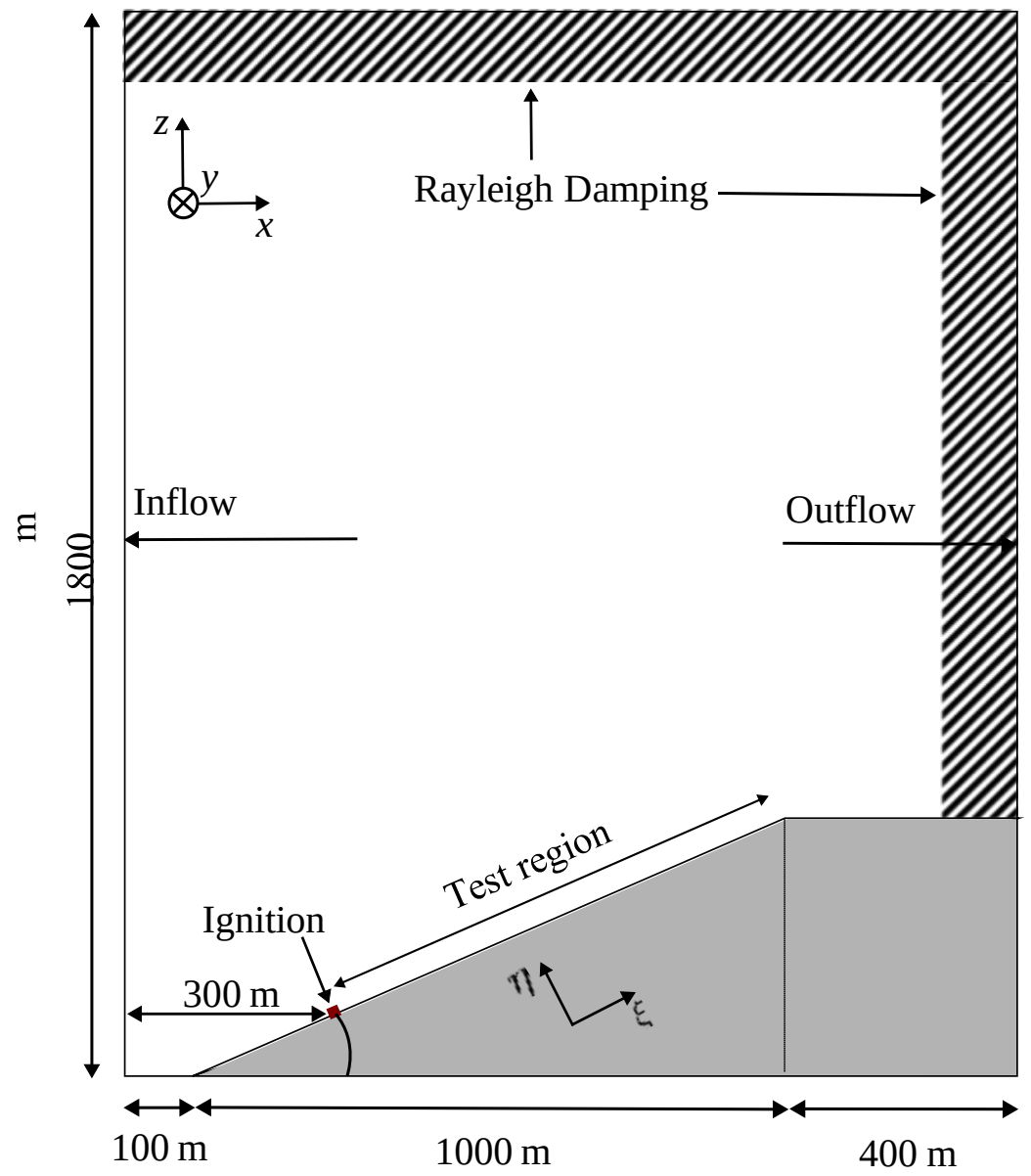

Figure 1: Schematic of a cross section of the computational domain for the ensemble simulations to examine effects of slope and wind on fire-spread behavior.

At the inflow, boundary conditions for velocity are prescribed from an atmospheric boundary-layer solution, which is computed prior to the ensemble simulations, following the procedure outline in Wang et al. (2023). In the present study, we consider a neutrally buoyant atmospheric boundary layer without heat transfer between the ground and the air. To accommodate different free stream wind speeds, we perform a precursor simulation with a mean wind speed at a height of 10 m above the ground of $U_{10,*} = 5$ m/s, where the subscript $*$ denotes the precursor boundary layer simulation. The precursor simulation is performed on the same computational domain as the fire simulation that has planar surface ($\alpha = 0°$). Periodic boundary conditions are applied along the streamwise direction (representing a temporally evolving boundary layer), with the flow driven by a Coriolis force. This precursor simulation is advanced for 30 flow-through times to obtain a statistically steady turbulent boundary layer. The flow-through time is defined as $\tau_{FTT} = L_x/U_{10}$ (where $L_x = 1500$ m is the streamwise extent of the domain). Subsequently, we collect the instantaneous velocity field at an $y - z$ plane in the middle of the domain, $u_*(y, z, t)$, over two flow-through times. This velocity field is then rescaled to match $U_{10}$ while keeping the turbulence intensity constant for all simulations. Mathematically, this rescaling procedure can be written as:

$$u_{in}(y, z, t) = \frac{U_{10}}{U_{10,*}} u_*(y, z, t), \tag{3.1}$$

where $u_*$ is the instantaneous velocity inflow field from the precursor simulations. Convective outflow boundary conditions are applied at the exit of the domain, free-slip conditions are employed at the top of the domain, and periodic boundary conditions are used in lateral directions. The terrain with constant slope is represented by an IB method, and Rayleigh-damping layers (Klemp & Lilly 1978) with a thickness of 5% of the domain size are applied at the top boundary and outflow to prevent gravitational waves and reverse flows in order to ensure stable boundary conditions.

In the present study, we consider the same fuel properties and ambient conditions (temperature, moisture) for all ensemble simulations. The fuel is represented by tall grass (fuel model 3 (Albini 1976)), which is distributed homogeneously, having a bulk-fuel density of 1.5 kg/m$^3$ and a fuel height of 1.5 m. The fuel moisture content is 14.2%. These conditions are representative of tall grass and the FireFlux-II fuel conditions. As such, these results are expected to be of broader relevance for prescribed burns and grassland fires (Cheney et al. 1993, Clements et al. 2014, 2019, 2007, Cruz et al. 2020, Paugam et al. 2021).

Figure 2 summarizes reported experiments on flat terrain ($\alpha = 0°$) for comparable fuel conditions (tall grass) in terms of the bulk fuel density $\rho_f$ and $U_{10}$ and rate-of-spread (ROS). The error bars in these figures represent the variation for each parameter in a series of experiments with the same mean condition. Based on the statistics, we see from Fig. 2a that the typical range of the bulk fuel density is between 0.5 to 3.5 kg/m$^3$, and $U_{10}$ ranges from 2 to 12 m/s. With this observation, we prescribed the bulk fuel density at 1.5 kg/m$^3$, which is representative of the conditions typically observed in grassland fires. To compare our simulation results with experiments, two additional simulations are performed under the condition of $U_{10} = 6\ \mathrm{m/s}$ without slope and bulk fuel densities being 0.4 and 2.5 kg/m$^3$. The wind speed $U_{10}$ is sampled between 2 and 10 m/s, with an increment of 1 m/s. In this study, we only consider the wind direction directly aligned with the slope direction. The slope angle $\alpha$ is independently varied between 0 and 30° to examine the effect of both slope and wind on the fire-spread behavior.

To assure that the flow field within the entire domain is fully developed, we advance the simulation for one flow-through-time prior to ignition. The fire is then ignited at $x = 300$ m across the entire lateral domain with a fire-line width of $l_x = 4$ m by setting the fuel temperature to $T_s = 600$ K.

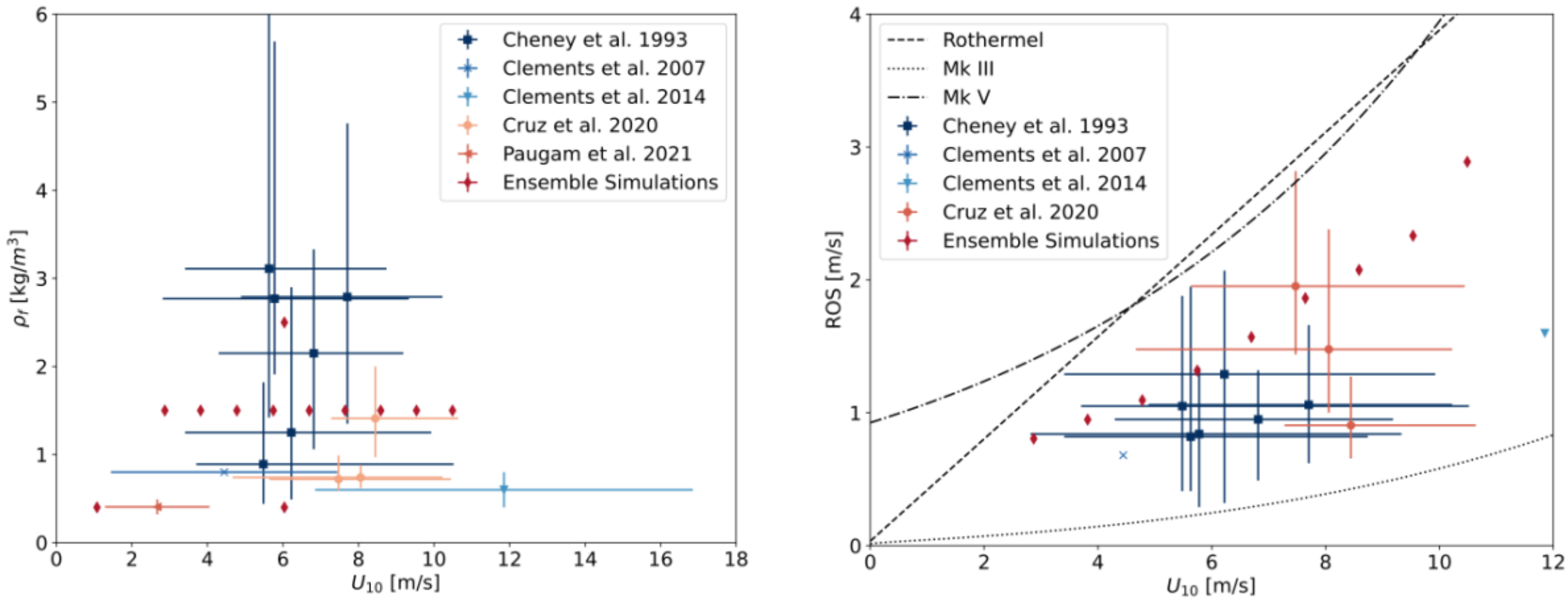

(a) $\rho_f - U_{10}$ (b) $R - U_{10}$

Figure 2: Comparison of operating conditions for ensemble simulations with observations (Cheney et al. 1993, Clements et al. 2014, 2007, Cruz et al. 2020, Paugam et al. 2021) for (a) fuel density $\rho_f$ and wind speed $U_{10}$, and (b) rate-of-spread ROS and wind speed $U_{10}$.

A total of 117 individual simulations were performed. Each simulation consists of 150 snapshots starting from the point of ignition, with a granularity of 1 s, which collectively provides 3D information pertaining to wind velocity, potential temperature, oxygen consumption, fuel distribution, and solid temperature. The simulations are labeled as U{XX}S{YY}, where XX is the wind speed $U_{10}$ and YY is the slope angle $\alpha$.

## Results

### *Instantaneous Flow-field Results*

To provide a qualitative comparison of the fire-spread behavior for the different conditions, in Fig. 3 we present instantaneous flow-field results for nine ensembles, corresponding to conditions $U_{10} = \{2, 6, 10\}$ m/s and $\alpha = \{0, 15, 30\}^\circ$ at a representative time instance where the fire front reaches the location of $x = 500$ m. The results in this figure show volume renderings of the potential temperature, and projections of axial ($x - z$ plane) and vertical ($y - z$ plane) velocity components. Comparing results along columns provides understanding about the effect of wind speed, showing that increasing wind speed results in a transition from plume-dominated to wind-dominated behavior. In contrast, for increasing slope angle the fire tends to attach closer to the surface, resulting in a reduction of flame height. This observation agrees with the literature, suggesting a transition between plume-dominated and wind-dominated fire modes as a function of wind speed and slope angle (Ju et al. 2019, Liu 2023).

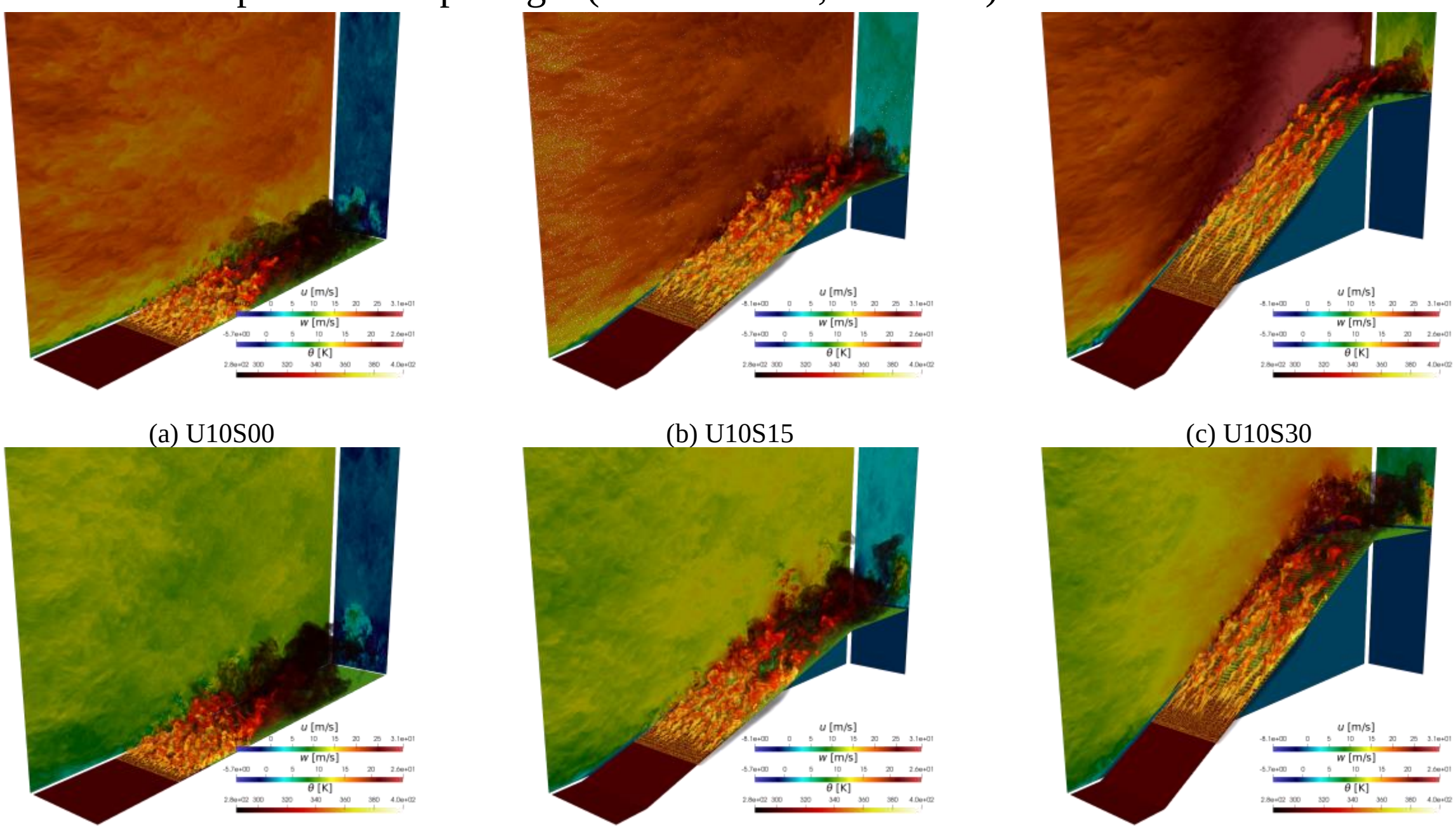

(a) U10S00 (b) U10S15 (c) U10S30

(d) U06S00 (e) U06S15 (f) U06S30

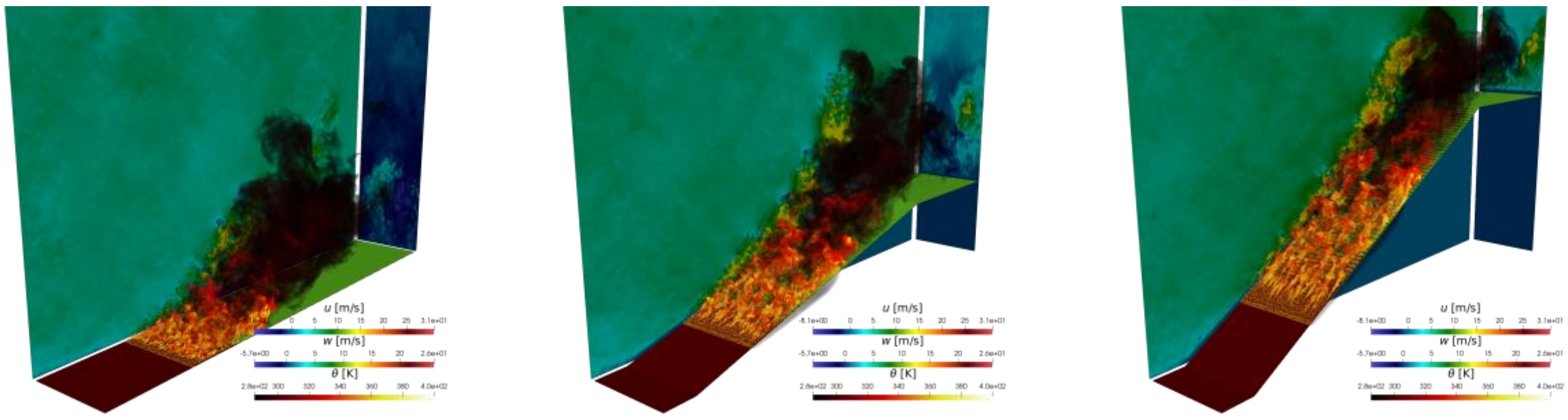

(g) U02S00 (h) U02S15 (i) U02S30

Figure 3: 3D visualizations of fire propagation for nine ensembles when the mean location of the fire front is x = 500 m: (a) U10S00, (b) U10S15, (c) U10S30, (d) U06S00, (e) U06S15, (f) U06S30, (g) U02S00, (h) U02S15, and (i) U02S30, showing instantaneous volume rendering of the potential temperature, the projection of axial velocity ($x - z$ plane) at $y = L_y/2$, and vertical velocity ($y - z$ plane) at $x = 1100$ m.

These results also provide insight about the coupled interaction of the slope and wind speed, clearly showing that these two parameters cannot be treated independently. To examine the coupling further, we extract the instantaneous fire front $\xi_f$ which is defined as the most upstream point with the potential temperature $\theta = 400$ K on the ground level. These results are presented in Fig. 4, showing isochrones of the fire front at discrete time intervals of 20 s. These results provide a direct illustration of the fire-front corrugation due to the coupling between buoyancy and convective heat-exchange processes (Finney et al. 2015) and fire intermittency by the atmospheric turbulent boundary layer (Wang et al. 2023). Specifically, it can be seen that increasing wind speed and slope not only result in a faster fire spread rate, but also in an increase in the amplitude and frequency of the fire-front corrugation.

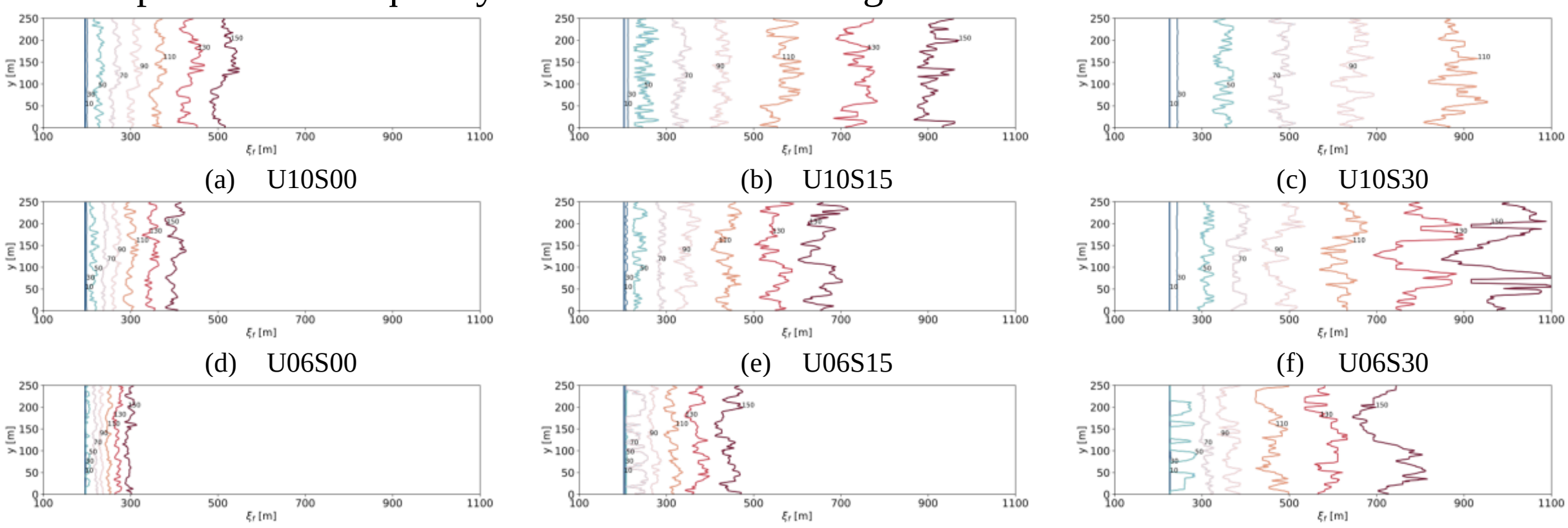

(a) U10S00 (b) U10S15 (c) U10S30 (d) U06S00 (e) U06S15 (f) U06S30 (g) U02S00 (h) U02S15 (i) U02S30

Figure 4: Isochrones of fire front locations $\xi_f$ for nine ensembles: (a) U10S00, (b) U10S15, (c) U10S30, (d) U06S00, (e) U06S15, (f) U06S30, (g) U02S00, (h) U02S15, and (i) U02S30. The fire front is defined as the most upstream point with $\theta = 400$ K.

*Statistical Analysis*

To examine the results from the ensemble simulation more quantitatively, we proceed with a statistical analysis. For this, we average the flow-field quantities along the homogeneous lateral direction, and compute a mean fire front $\langle \xi_f \rangle_y = L_y^{-1} \int \xi_f dy$ and corresponding fluctuation of the fire front $\xi_f' = \xi_f - \langle \xi_f \rangle_y$. The laterally averaged instantaneous ROS $\langle R \rangle_y$ is then computed as:

$$\langle R \rangle_y(t) = \frac{d\langle \xi_f \rangle_y}{dt}. \tag{4.1}$$

Results from this analysis are presented in Fig. 5, which correspond to the nine cases shown in Figs. 3 and 4. The mean fire location and ROS are shown by solid lines, with the shaded area and error bars representing the instantaneous fluctuation of the fire front location and ROS. The results in Fig. 5 show that the fire dynamics can be characterized by three distinct phases. The different phases are more pronounced for fires with larger slope angles. The first phase is associated with the fire transition following the fire-line ignition. In this phase, the fire advances only slowly without significant fire corrugation (see Fig. 4). Following this transition phase, the fire advances with a nearly constant acceleration that is marked by sporadic intermittency. The spatial extent of this quasi-equilibrium phase depends on the wind speed and slope angle and reduces with increasing $\alpha$ and $U_{10}$. Based on results shown in Fig. 5, we identify this phase as the region where the fire front location is between $\xi_0 = 400$ m to $\xi_1 = 800$ m. The dashed lines in Fig. 5(d-f) show linear fits of the ROS with respect to time,

$$\langle R \rangle_y^*(t) = \langle R \rangle_{y,0} + a(t - t_0), \qquad \text{for } t \in [t_0, t_1], \tag{4.2}$$

where $t_0$ and $t_1$ are the times when the mean fire front reaches $\xi_0$ and $\xi_1$, respectively, $\langle R \rangle_{y,0}$ is the ROS at $t_0$, $a$ is the linear acceleration of the fire, and the superscript $*$ denotes the quasi-equilibrium period. By integrating Eq. (4.2), we obtained the mean ROS in the quasi-equilibrium phase as:

$$\overline{\langle R \rangle_y^*} = \frac{1}{t_1 - t_0} \int_{t_0}^{t_1} \langle R \rangle_y^*(t) dt = \langle R \rangle_{y,0} + \frac{a}{2}(t_0 + t_1), \tag{4.3}$$

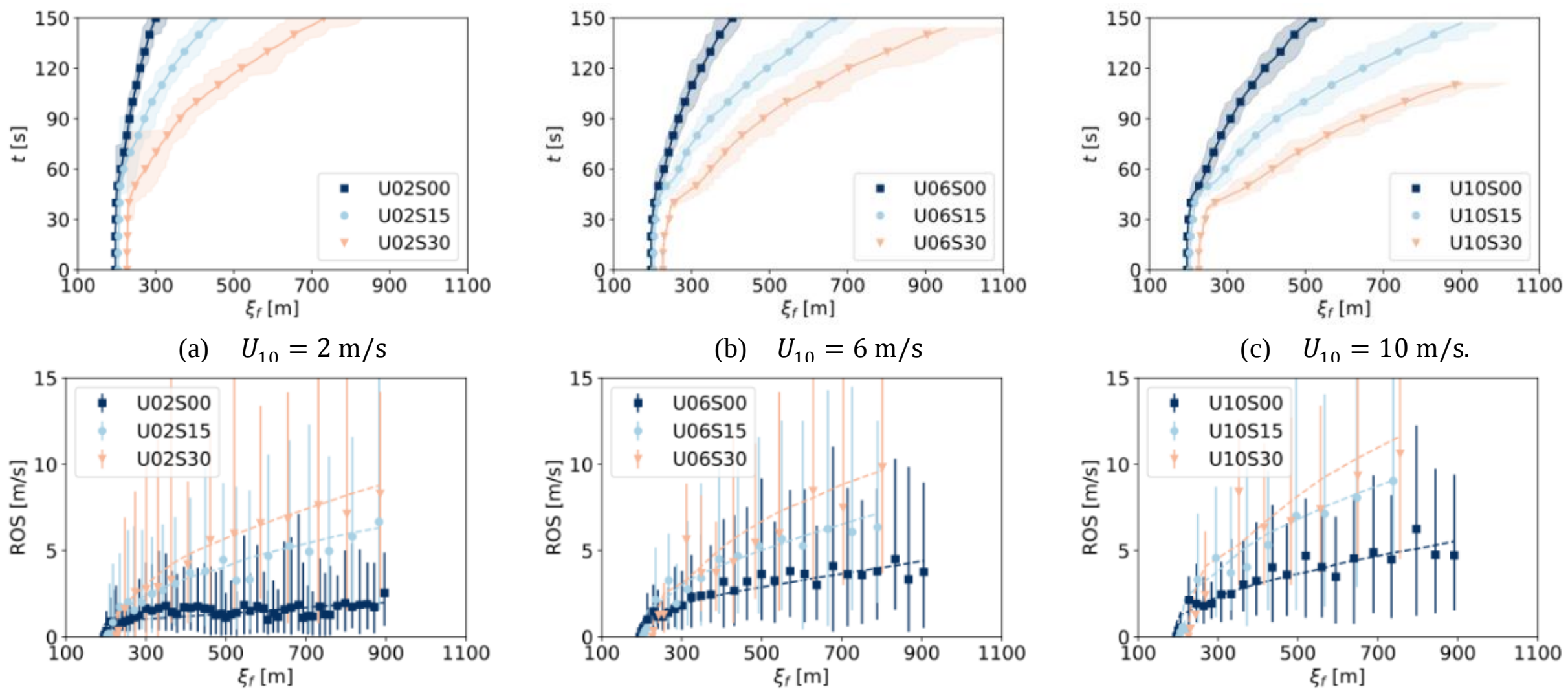

(a) $U_{10} = 2$ m/s (b) $U_{10} = 6$ m/s (c) $U_{10} = 10$ m/s.

(d) $U_{10} = 2$ m/s (e) $U_{10} = 6$ m/s (f) $U_{10} = 10$ m/s

Figure 5: Comparison of the statistical results for (top) mean fire-front location $\langle \xi_f \rangle_y$ and (bottom) rate-of-spread ROS $\langle R \rangle_y$ for three different wind speeds (a,d) $U_{10}$ = 2 m/s, (b,e) $U_{10}$ = 6 m/s, and (c,f) $U_{10}$ = 10 m/s and three slope angles ($\alpha$ = {0,15,30}°) shown by different colors in each panel.

The mean ROS and linear acceleration in the quasi-equilibrium period are presented in Fig. 6. From these results, it can be seen that the fire spreads at almost a constant rate at low wind speeds and small slope angles. This is consistent with observations and several empirical models (Sullivan 2009), where wind speed and slope angle are considered as independent parameters, not considering feedback on the fire-spread rate. Comparisons of predicted zero-slope mean ROS with measurements and the phenomenological models by Rothermel and McArthur's fire danger index (Mk V) (Noble et al. 1980, Rothermel 1972) are presented in Fig. 2b. Our predictions show overall good agreement with observations and the empirical fire-spread models capture the trends, with Rothermel's model providing a slightly better agreement. All results fall within the range of the experimental observations, while showing a slightly faster trend, which we attribute to the lower bulk-fuel density.

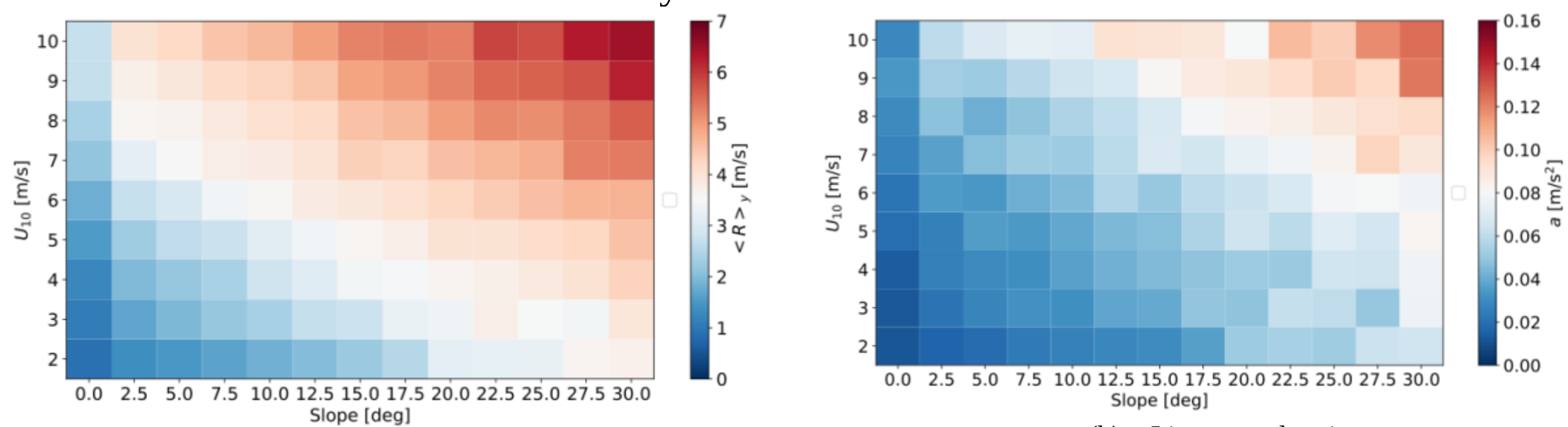

(a) Mean ROS

(b) Linear acceleration

Figure 6: Quantitative analysis of ensemble simulations as a function of slope and wind speed, showing (a) mean ROS, $\langle R \rangle_y$, and (b) linear acceleration $a$, computed from Eq. (4.2) during the quasi-equilibrium phase for $\xi \in [400{,}800]$ m.

Compared to the zero-slope results, Fig. 6b shows that with increasing slope angle the unsteady behavior becomes more pronounced. This is attributed to the coupling of the fire to the atmosphere by enhanced air entrainment, buoyancy, and heat transfer to the fuel bed upstream of the fire front. The general trend agrees with observations by Cheney & Gould (1997). In particular, in an experiment with a wind speed of 4.5 m/s, the observed average acceleration is close to 0.04 m/s$^2$ (Cheney & Gould 1997), which is close to $a = 0.02\ \mathrm{m/s^2}$ from the simulations. Note that the acceleration is underpredicted by the simulation in this case. This may be due to the underrepresentation of the wind gusts that drive the change of wind directions, which has been identified as a key factor for fire acceleration.

As the fire approaches the plateau at the end of the slope, the fire growth increases rapidly due the development of the recirculating flow field on the plateau, which can be seen from the instantaneous streamwise velocity fields in Fig. 3. Note that the fluctuations and corrugations of the fire front increase with both slope angle and wind speed. However, Fig. 6b shows that as the wind speed increases, the dependence in fire growth on the slope angle diminishes, suggesting that the wind speed is the dominating factor for the fire acceleration in the present configurations.

We proceed by quantifying the fireline intensity and fuel consumption rate. These quantities are important properties of line fires, measuring the heat release rate of wildfires with direct implication in fire management (Alexander 1982). For this, we compute the fireline intensity as (Finney et al. 2021):

$$I_B = H_c m_c'' \langle R \rangle_y, \tag{4.4}$$

where $H_c$ is the heat of combustion and $m_c''$ is the fuel consumption per unit area in the flame zone, which is determined as:

$$m_c'' = -\frac{1}{\dot{A}_{burn}} \int_\Omega \dot{\omega}_f d\Omega, \tag{4.5}$$

where $\Omega$ is the simulation domain, $\dot{A}_{burn}$ is the rate of change of the burned area, and $\dot{\omega}_f$ is the rate of fuel consumption. With the burn area being determined as the area where the bulk fuel density is below 20% of its initial value, its rate of change at time $t_n$ is computed as $\dot{A}_{burn}^{t_n} = (\dot{A}_{burn}^{t_{n+1}} - \dot{A}_{burn}^{t_{n-1}})/2\Delta t$. Figure 7a shows the heat density $H_c m_c''$, computed from the simulations in units of MJ/m². The mean heat density for the present study is found to be 25.9 MJ/m², with a standard deviation of 3.3 MJ/m² and a maximum difference of 13.1 MJ/m² for all simulations. These results show that the heat density only exhibits a weak dependence on wind speed and slope angle, which is in agrees with observations by (Finney et al. 2021).

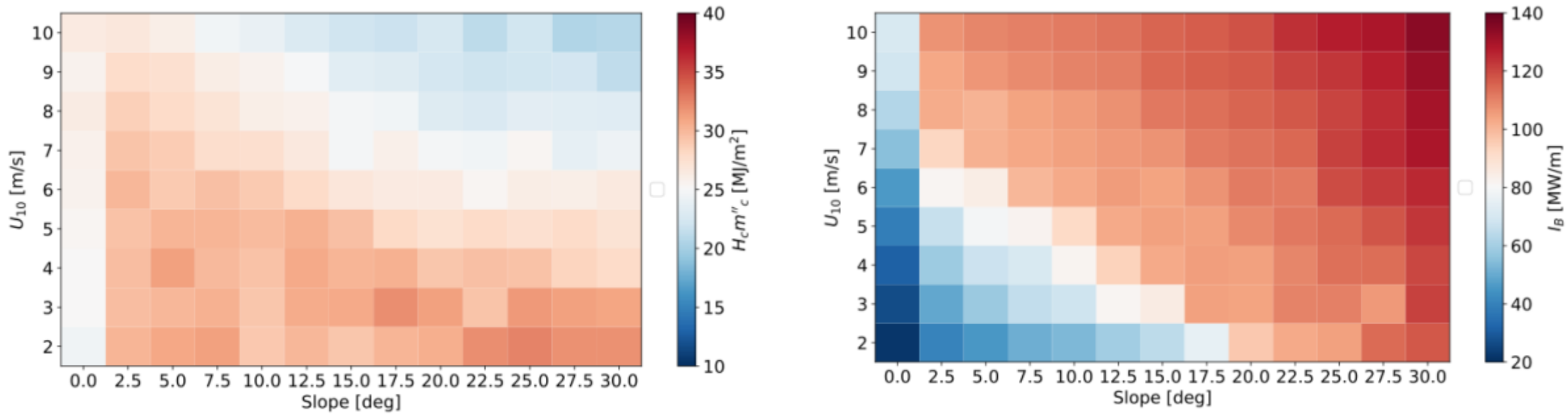

(a) Heat density (b) Fire intensity

Figure 7: Quantitative analysis of ensemble simulations as a function of slope angle and wind speed, showing (a) Heat density, $H_c m_c''$, and (b) Fireline intensity, $I_B$.

In contrast to the heat density, the fireline intensity shows a strong dependency on the wind speed and slope angle, as shown in Fig. 7b. Given that the fireline intensity is defined as the product between the heat density and ROS, and the heat density stays relatively constant with respect to the wind and slope angle, the fireline intensity shows the same response with the wind and slope as the ROS. With increasing wind speed and slope angle, the fireline intensity is larger. Similar to the ROS, shown in Fig. 6a, the dependency on the wind and slope factors is nonlinear, which is indicated by the relatively faster growth rate with respect to the slope angle for the cases with a faster ambient wind.

*Wind-slope Effects on Fire-spread Rate*

To examine the effect of the wind and slope conditions on the fire growth, we compare ROS obtained from the simulations with the Rothermel model following Rothermel's original definition and Nelson's modifications. The original definition of the Rothermel model considers a linear combination of the wind and slope factors,

$$R(u, \alpha) = R_0[1 + \phi_w(u) + \phi_s(\alpha)], \tag{4.6}$$

where $R_0$ is the zero-wind, zero-slope ROS, $u = U_{10}$, and $\alpha$ is the slope of the fuel bed. The modification due to Nelson (2002) to the model adjusts the effective wind speed geometrically by incorporating the slope angle. The vertical velocity due to buoyancy is computed as

$$w = \left(\frac{2 g I_B}{\rho_a c_p T_a}\right)^{1/3}, \tag{4.7}$$

where $I_B$ is computed with the assumption that $m''_c$ is the total available fuel loading. With this, the effective wind speed parallel to the slope, assuming that the wind is aligned with the gradient of the slope, is computed as

$$u^\star = U_{10} \cos \alpha + w \sin \alpha. \tag{4.8}$$

The ROS is then computed following Rothermel's formulation with the slope factor omitted:

$$R(u^\star) = R_0[1 + \phi_w(u^\star)]. \tag{4.9}$$

Equations (4.7) to (4.9) are solved using Newton iteration to obtain the ROS.

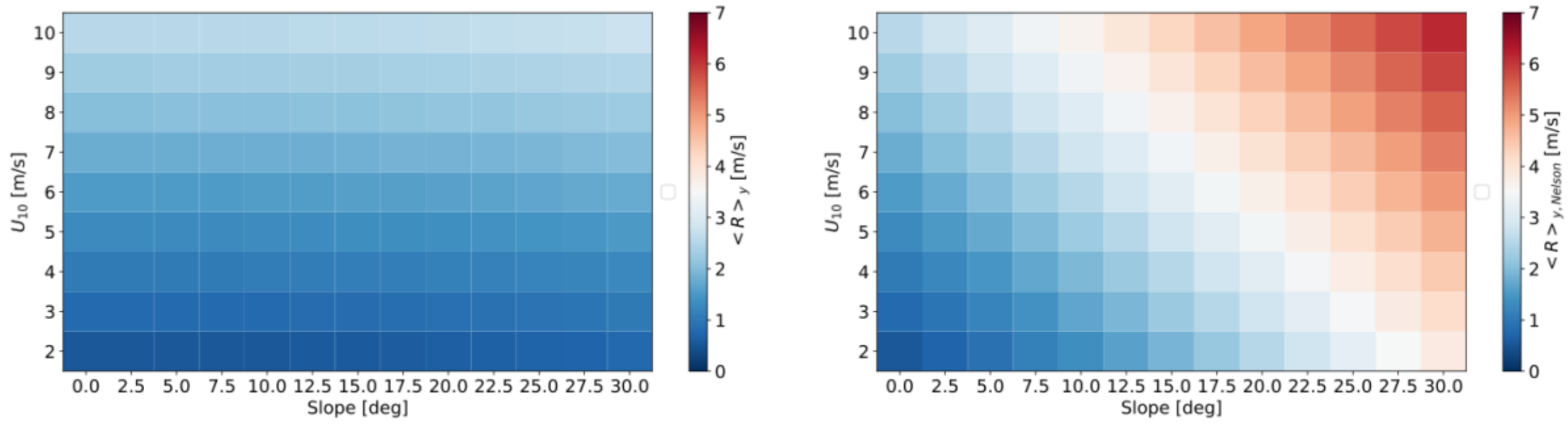

(a) Rothermel's model (b) Nelson's modified model

Figure 8: Comparison of rate-of-spread prediction from (a) Rothermel model and (b) Nelson's modified model.

Figure 8 shows results over the parameter space corresponding to that of the ensemble simulation. The Rothermel model with Nelson's modification shows that the ROS has a stronger dependency on the slope compared to the original formulation. Different from the original Rothermel model, in which the slope and wind factors are derived from conditions in the absence of wind and slope, respectively, the modified Rothermel model considers the non-linear coupling of the slope on the wind speed. This coupling effect is shown to be important from the simulation results. Good qualitative agreement is observed between the modified Rothermel model and the simulation results. Figure 9 shows the mean ROS for the modified wind speed $u^\star$. $u^\star$ of the simulation is computed with Eqs. (4.7) and (4.8) for direct comparisons with Nelson's analysis. A positive correlation between these two factors with a relatively small variance is observed, suggesting that the coupling between the buoyancy-induced vertical velocity and the wind speed is important for predicting the ROS on a slope. Note that the mean ROS obtained from the simulations is higher than what is predicted by Nelson's model for the range of $u^\star$ considered in this study. This may be caused by the change in wind directions due to turbulence that is induced from the fire/atmosphere interactions, which needs further investigations.

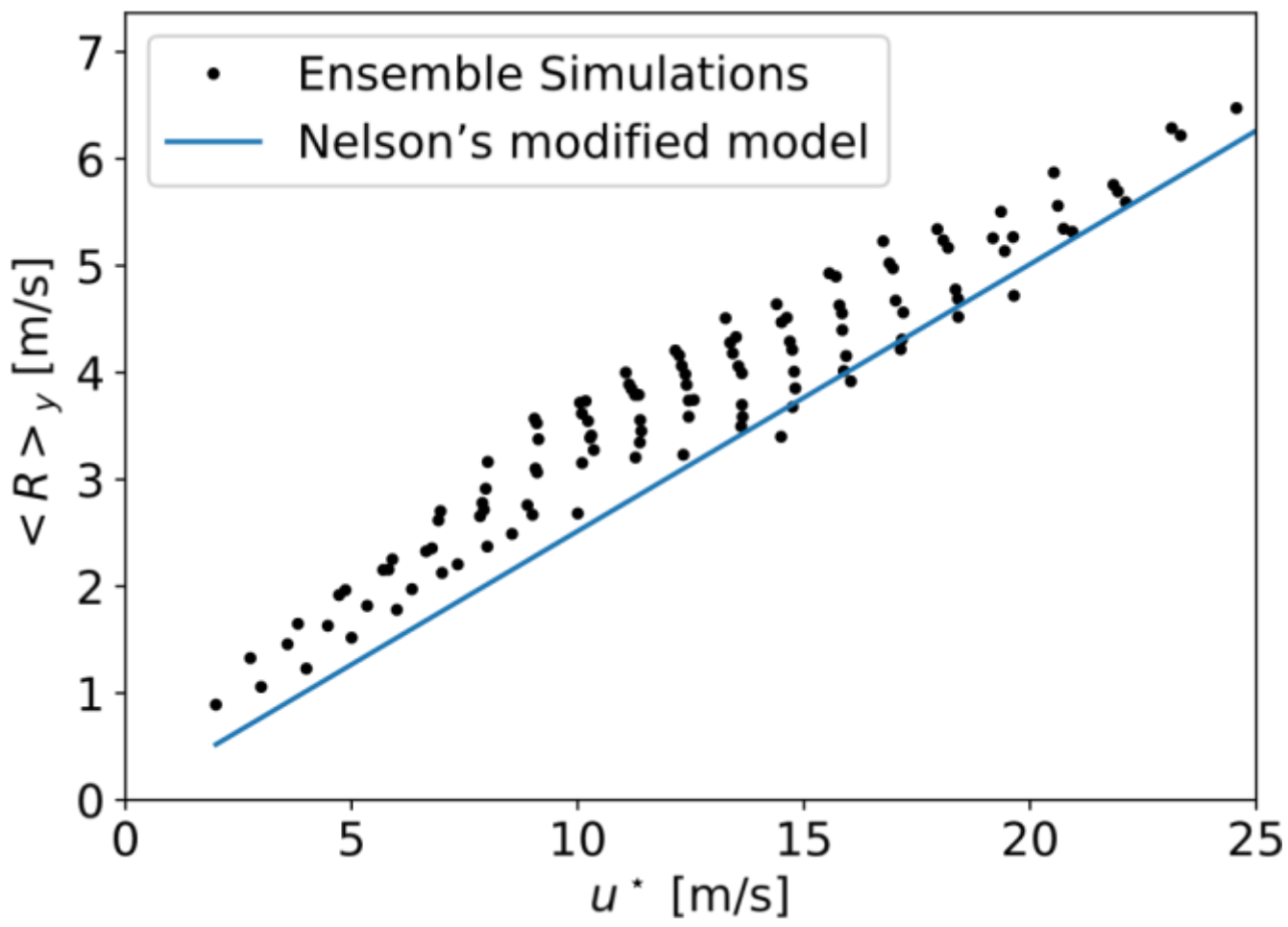

Figure 9: ROS from the simulation data for the effective wind speed computed with Eq. (4.8)

To further investigate the fire dynamics and effects of buoyancy, we compute the convective Froude number (Sullivan 2007)

$$\mathrm{Fr} = 1.29(u - R)\left(\frac{\rho_a c_p T_a}{g I_B}\right)^{1/3}, \tag{4.10}$$

where $u$ is computed from Eq. (4.8) to incorporate the impact of both the wind speed and slope. Figure 10 shows the Froude number parameterized by the wind speed and slope. Literature identified that the Froude number can be used as parameter to quantify the transition between a plume-dominated and a wind-dominated fire (Moinuddin et al. 2018, Sullivan 2007). The critical Froude number for this transition is found to be around 0.5 (Moinuddin et al. 2018, Morvan & Frangieh 2018), which corresponds to the lower-left region in Fig. 10, with slopes smaller than 15° and wind speed less than 7 m/s.

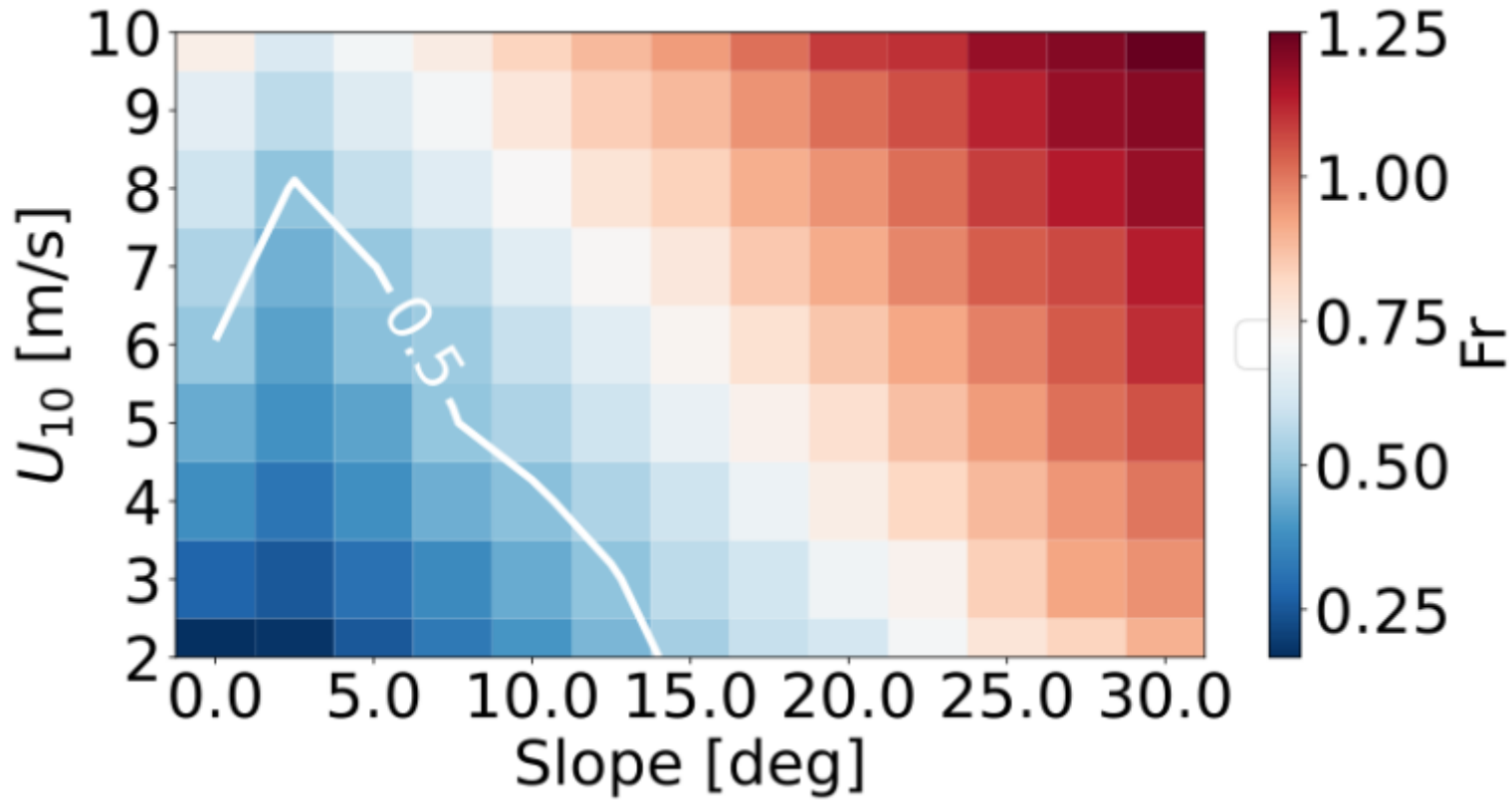

Figure 10: Froude number computed from Eq. (4.10).

The transition of the fire spread from the plume-dominated mode to the wind-dominated mode is demonstrated with the mean potential temperature and streamlines near the fire front in Fig. 11, shown in the slope-oriented coordinate system. The origin of the slope-aligned coordinate $\xi$ is shifted to the location of the fire front before computing the mean across the $y$-direction. The mean ambient wind speed is removed from the velocity field to improve the illustration of the local interaction of the fire and the flow field. The three cases selected are associated with $U_{10}$ = 4 m/s, with slope angles being 0°, 10°, and 30° that correspond to Fr < 0.5, ∼ 0.5, and > 0.5, respectively. In Fig. 11a we see that the fires are almost vertically oriented, which indicates that the buoyancy is dominating the convection from the upstream wind. The large-scale fluid motion is perturbed by the fire. Additionally, air entrains from upstream of the fire, which reduces the convective heat flux from the flame to the unburned fuel bed. In contrast, in the case with Fr > 0.5, as shown in Fig. 11c, the flame attaches close to the ground, and the streamlines aligned along the vertical direction without large-scale perturbations. As a consequence, the unburned fuel is preheated mainly by the convective heat flux from the flame. When Fr∼0.5, as in Fig. 11b, the flame leans closer towards the ground compared to Fr < 0.5 in Fig. 11a, and the flow field is less chaotic with weakened upstream air entrainment. The mode of unburned fuel preheat starts to transition from radiation dominated to wind dominated.

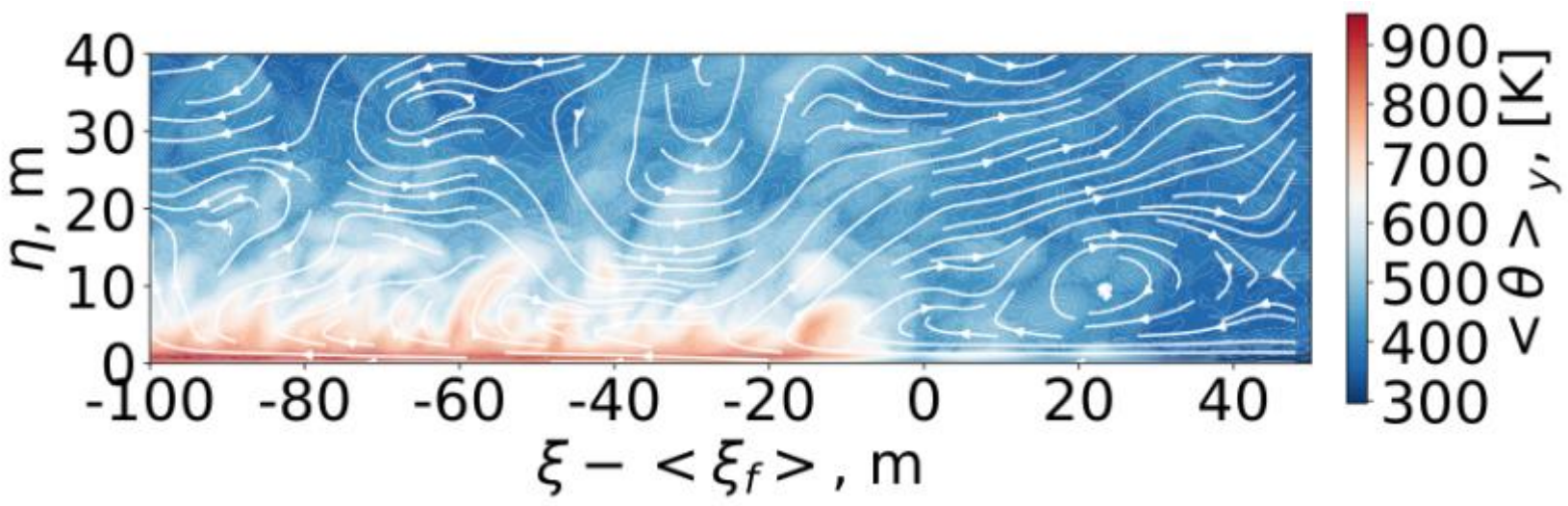

(a) U04S00.

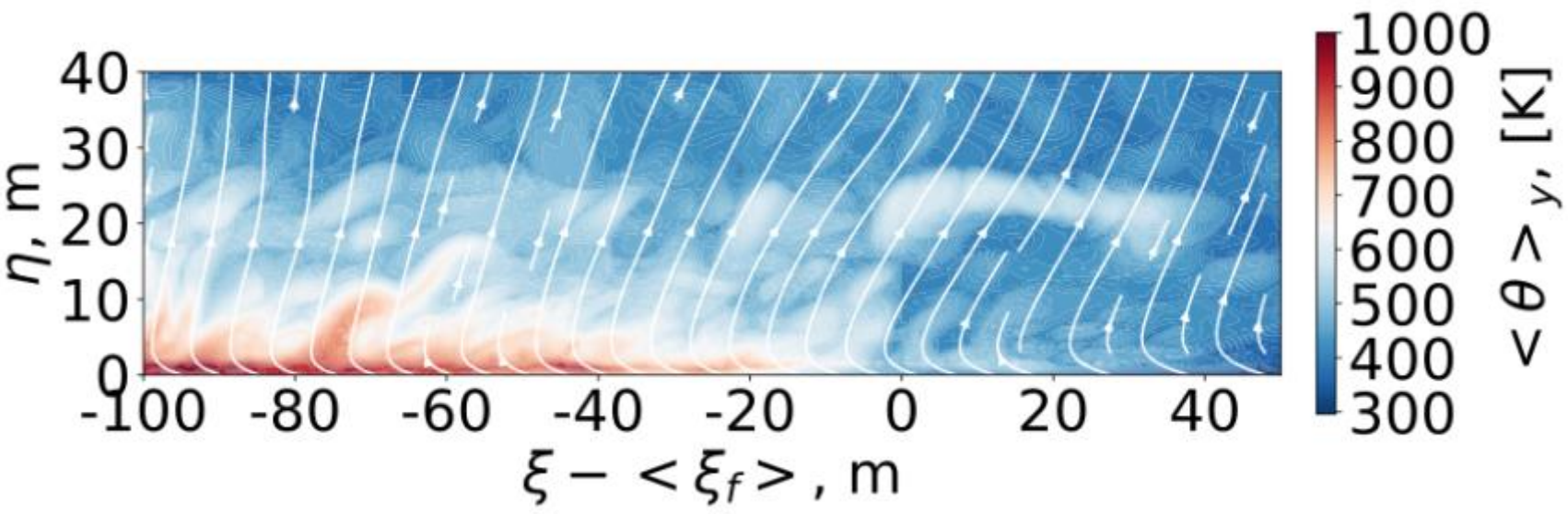

(b) U04S10.

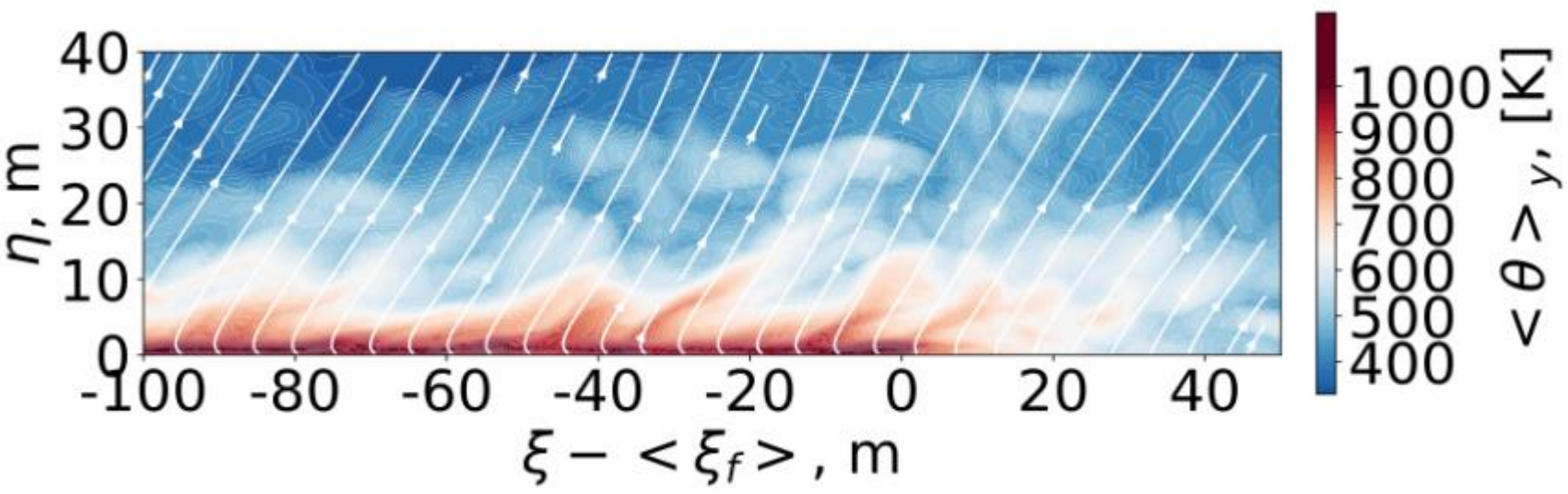

(c) U04S30.

Figure 11: Comparison of the potential temperature and flow field at the fire front from (a) U04S00, (b) U04S10, and (c) U04S30.

We find that this plume-dominated regime correlates with the regime where large differences in ROS are found between the simulations and the modified Rothermel model. This is due to the strong coupling between the fire dynamics and the fire-induced turbulence in a plume-dominated fire, which requires more investigations.

**Conclusions**

In this study we introduce a new platform for high-fidelity ensemble wildfire simulations using the large-eddy simulation SWIRL-FIRE framework, which is designed and developed for running high-resolution three-dimensional fire simulations on tensor-processing units (TPU) with high computational efficiency. The ensemble simulation framework couples SWIRL-FIRE with the hyperparameter optimization platform VIZIER for automated parameter sampling, simulation scheduling and execution, post-processing analysis, and data management.

To demonstrate this ensemble simulation platform, we examined the couple of wind-slope effects, which has been an outstanding issue due the experimental limitations of prescribed and laboratory fires. The simulations are performed based on a grass-land line-fire configuration. The wind speed parameter is defined by the mean streamwise velocity at 10 m above the ground level ($U_{10}$) and is sampled between 2 m/s and 10 m/s with a 1 m/s interval. The angle of the slope is sampled between 0$^\circ$ and 30$^\circ$ with a 2.5$^\circ$ interval. A domain of size 1500 × 250 × 1800 $m^3$ is discretized with a uniform mesh of 1 × 1 × 0.5 $m^3$ resolution to capture the fire-atmosphere interaction. We find good agreement of the fire ROS between simulations with 0$^\circ$ slope and experimental observations that are performed under similar conditions.

Results from the ensemble simulations show a strong coupling effect of the wind speed and slope on the fire propagation and intermittency. As the increase in wind speed and slope angle promote the ROS and fire front corrugation, the contribution by these two factors are not linearly superimposed. This observation is evident from the comparison of the simulation results and the Rothermel model. In Rothermel's original formulation, the wind and slope factors are combined linearly. In a version of the model revised by Nelson, the wind factor incorporates the vertical velocity induced by buoyancy, which considers a non-linear coupling between the wind speed and slope angle. Results from the ensemble simulations show good agreements with the modified Rothermel model. Results from the Rothermel model with the original formulation show a maximum of 50% difference from the other two approaches at high wind speed and large slope angles.

We delineate regimes between wind-dominated fires and plume-dominated fires parameterized by the convective Froude number that is formulated as a function of wind speed and slope angle. Our results show that for Froude numbers below 0.5, the fire spread is dominated by buoyancy with a visible plume formed by the upward motion. As the convective Froude number increases above 0.5, the fire plume gets closer to the ground, which suggests that convection is dominating the fire spread.

We note that a relatively large difference in ROS is observed between results obtained from the simulation and the fire spread model in the plume-dominated regime, which is due to the complex interaction between fire-induced turbulence and atmospheric dynamics. Investigating and improving the modelling capability is beyond the scope of this paper, but it is an interesting area of study with the ensemble simulation framework.

The FIREBENCH dataset is available at https://console.cloud.google.com/storage/browser/firebench for further analysis, for evaluating fire-spread prediction models, and for training and benchmarking ML models.

## Acknowledgments

We thank Tyler Russell and Carla Bromberg for their support on releasing the dataset on the Google cloud platform.

**Appendix: Comparisons against experimental data**

To assess the accuracy of the SWIRL-FIRE solver in predicting the wildfire behavior for conditions with sloped terrain and different vegetation, we perform simulations of the experiments conducted by Morandini & Silvani (2010). The simulations are performed using the same domain size, mesh resolution, and boundary conditions as for the ensemble simulations that were shown in Fig. 1. To cover a range of representative conditions, we consider four experimental cases that are characterized by different slope angles, wind speeds and wind directions, fuel loadings, and fuel moisture contents. The experimental conditions are summarized in Table A1.

Table A1: Summary of the experimental conditions (Morandini & Silvani 2010) that are used for the comparisons of the simulation results. The last row summarizes ROS predictions with $R^2$coefficient of determination.

| **Experimental case** | **1** | **2** | **3** | **5** |
|---|---|---|---|---|
| Slope [∘] | 20 | 17 | 26 | 26 |
| Wind velocity [m/s] | 3.3 | 3.6 | 2.0 | 3.4 |
| Wind directions [∘] | 67 | 131 | 54 | 176 |
| Fuel load [kg/m$^2$] | 5.4 | 7.4 | 4.1 | 1.4 |
| Fuel moisture content [%] | 89 | 62 | 100 | 51 |
| Fuel height [m] | 0.8 | 1.2 | 0.7 | 0.3 |
| Experimental ROS [m/s] ($R^2$) | 0.18 (0.91) | 0.29 (0.86) | 0.20 (0.92) | 0.35 (0.96) |
| Predicted ROS [m/s] ($R^2$) | 0.17(0.99) | 0.25 (0.98) | 0.20 (0.99) | 0.37 (0.97) |

In these simulations, a line fire is ignited at 250 s after the cold flow reaches steady-state conditions, after which the fire is advanced for another 150 s. The ignition lines for all validation simulations are placed at $x = 500$ m. The length of the ignition lines were set to be the width of the plot for each experiment. Wind direction, moisture content, fuel density, and fuel height in these simulations are prescribed conditions reported in the experiments and summarized in Table A1, assuming homogeneous properties.

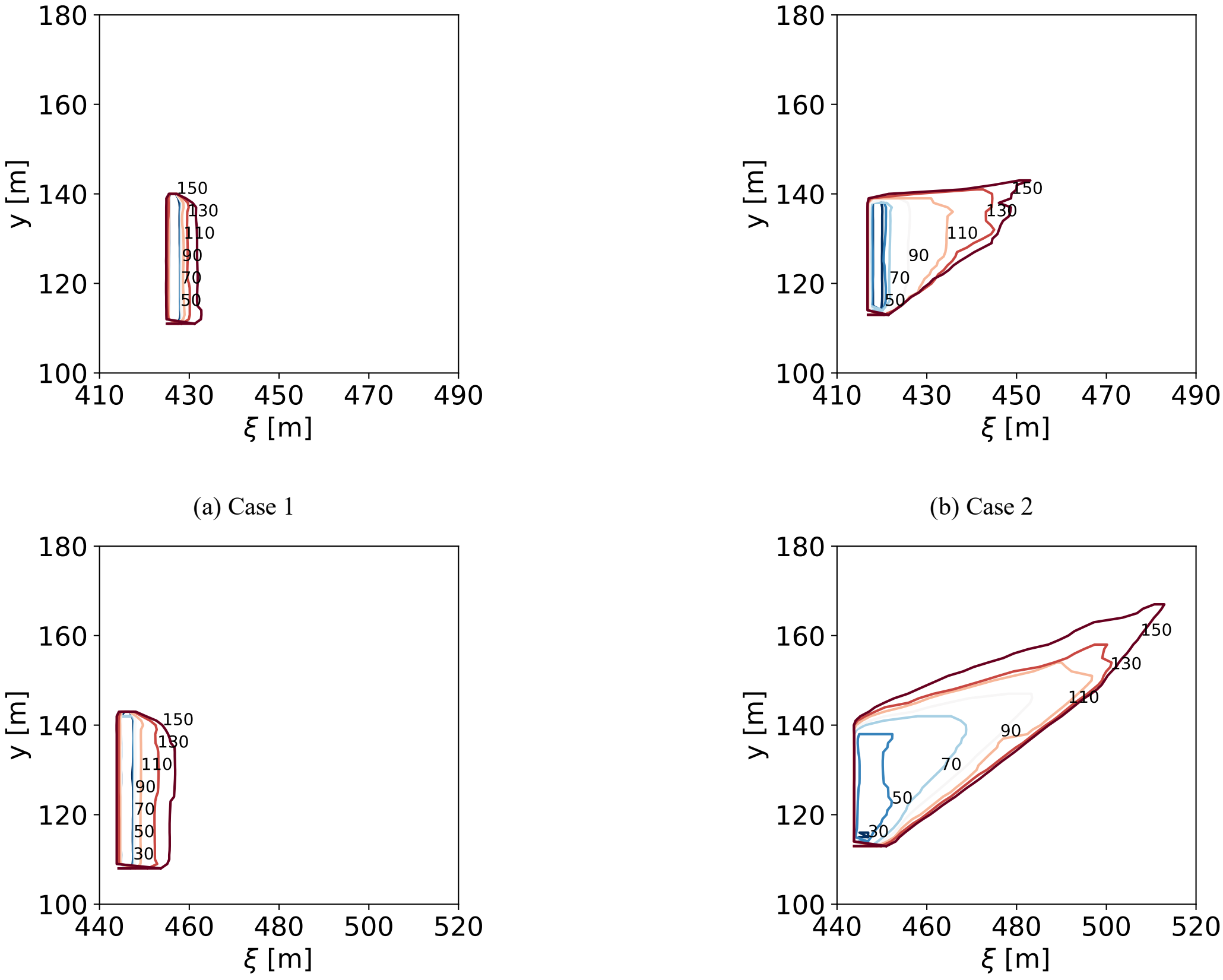

(a) Case 1

(b) Case 2

(c) Case 3

(d) Case 5

Figure A1: Isochrones for fire arrival times. Contours are evaluated for the gas-phase temperature of 400°C at 20 s intervals for (a) Case 1, (b) Case 2, (c) Case 3, and (d) Case 5 from Morandini & Silvani (2010) and reported in Table A1.

Results of computed isochrones for the four cases are presented in Fig. 1A, showing the effects of wind direction, slope angle, and fuel density and moisture on the fire-spread dynamics. The fire front was computed from the most upstream location of the fire front and is shown in Fig. A2. Following the procedure, outlined in the main text (Section “Statistical Analysis”), we evaluate the mean ROS from the fire-front evolution during the quasi-equilibrium phase and marked by the line in light color in Fig. A2. The so computed mean ROS for the cases 1, 2, 3, and 5 are reported in the last row of Table A1 where they are compared with experimentally reported data. Overall, it can be observed the simulation captures the experimental trends of mean ROS with response to slope angle, wind, fuel density, and fuel moisture content.

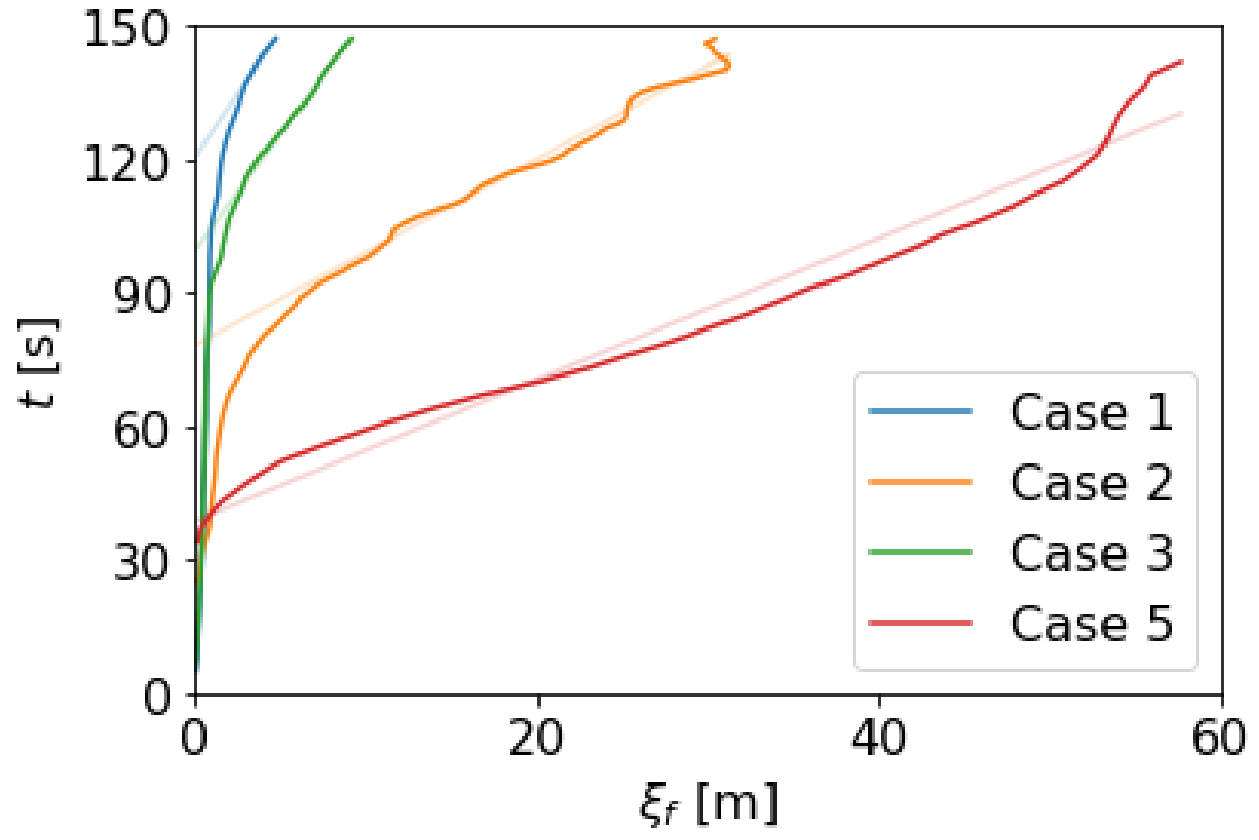

Figure A2: Predictions of instantaneous fire front location for the four validation cases 1, 2, 3, 5 from Morandini & Silvani (2010); Regression lines, shown in light colors, are used to compute the mean ROS, reported in Table A1.